\documentclass[aps,pra,superscriptaddress,amsmath,amsfonts,amssymb,floatfix,nofootinbib]{revtex4}
\usepackage{enumerate}
\usepackage{mathtools}
\usepackage{amsmath}
\usepackage{graphicx}
\usepackage{epsfig}
\usepackage{sidecap}
\usepackage{hyperref,stackrel}
\usepackage[normalem]{ulem}
\usepackage{color}
\usepackage{enumerate}
\usepackage{bm}
\usepackage[utf8]{inputenc}
\usepackage{amsmath}
\usepackage{amsfonts}
\usepackage{amssymb}
\usepackage{graphicx}
\usepackage{enumitem}
\usepackage{physics}

\begin{document}

\title{Quantum Teleportation using Quantum Candies}
\author{Nikhitha Nunavath} \thanks{nikhithanunavath@students.iisertirupati.ac.in}
\affiliation{Jaypee Institute of Information Technology, A-10, Sector-62, Noida, UP-201309, India}
\author{Sandeep Mishra} \thanks{sandeep.mtec@gmailcom}
\affiliation{Jaypee Institute of Information Technology, A-10, Sector-62, Noida, UP-201309, India}
\author{Anirban Pathak} \thanks{anirban.pathak@gmail.com}
\affiliation{Jaypee Institute of Information Technology, A-10, Sector-62, Noida, UP-201309, India}

\begin{abstract}
Quantum Candies or Qandies provide us with a lucid way of understanding the concepts of quantum information and quantum science in the language of candies. The critical idea of qandies is intuitively depicting quantum science to the general public, making sense as most of the research in this domain is funded by the taxpayers. The qandies model is already used to explain the essential concepts of quantum science and quantum cryptography. However, teleportation and related concepts are yet to be explained. Motivated by this fact, we investigate and extend the idea of Jacobs and Lin-Mor-Shapira to explain teleportation using qandies. Here, we explicitly design the teleportation protocol and perform a circuit model using qandy gates. The protocol is successful when the correlated qandies are appropriately pre-shared and use of some local operations at both ends. The model we develop can be a valuable tool for science and engineering educators who want to help the general public to gain more insights into quantum science and technology. 

\end{abstract}
\maketitle

\section{Introduction}

There are many aspects of science and technology education. Specially, when it's related to science and technology teaching or communication as it demands different tools and methodology for different level of learners/participants. For example, interest on black hole or big bang theory is not restricted to professional physicists. Common people are also interested about them. However, to explain the essential ideas of black hole or big bang to (a) a set of physicists (b) a set of schools students who has science as a subject and (c) a set of citizens who has not studied science after their school, we must follow different strategies. Same is true for all the existing and upcoming technologies. Moreover, scientists  frequently write for their peers and publish in the journals focused on publishing advanced research, but many people who are not scientists or work as scientists in other domains are also interested to know about the recent developments in specific aspects of science and technology. Recently, such an interest is observed in the domain of quantum technology as we are approaching second quantum revolution \cite{dowling2003quantum,ladd2010quantum,pirandola2020advances,mannalatha2023comprehensive} and several articles are appearing in newspapers and magazines. Moreover, a great amount of interest is observed to be towards understanding the phenomena which occur in quantum world, but have no analogue in classical world. One such phenomenon is quantum teleportation \cite{bennett1993teleporting,hu2023progress}. In quantum teleportation, information about a system is communicated from one place to other place, but after following the steps of the protocol, the information gets destroyed from the sender's end and recreated at the receiver's end. Moreover, in the process  the information is never found in the channels that connect the receiver and the sender. There is no such process in the classical world. One may note that sending a letter from a city to other city is not teleportation as the letter travels through the channel joining the cities. Even sending a fax is not teleportation as the original remains with the sender while the receiver receives a copy. Interestingly, the laws of physics prohibit such copying (cloning) in the quantum world \cite{wootters1982single}.  Here, we aim to illustrate the concept of quantum teleportation without using any kind of mathematics. Instead, we will highlight the concept via use of specific type of candies which we would refer to as quantum candies or qandies. This will be in line with some of the existing works where qandies have been used to explain various other concepts of quantum information \cite{lin2020quantum,lin2021illustrating,mor2022digital}. 

Before we illustrate quantum teleportation with qandies, we would like to note that ever since its inception, counter-intuitive nature of quantum mechanics has often amazed and puzzled scientists. In the early days of quantum mechanics, nonclassical features of quantum mechanics were of interest to a set of scientists only. However, the interest on so called weird properties of quantum mechanics, like nonlocality, contextuality, entanglement, etc., have been enhanced in the recent past with the advent of quantum computing and quantum communication. There are some obvious reasons behind the recent enhancement of interest on these aspects of quantum mechanics. Firstly, several of the counter intuitive properties of quantum mechanics have now been exploited to develop devices leading to quantum technology. Specifically, quantum random number generators and quantum key distribution systems are now available commercially \cite{pirandola2020advances,mannalatha2023comprehensive}. Secondly, the idea of quantum computing introduced by Feynman \cite{feynman1986quantum} has also been developed over time. Companies like IBM, Google, etc. are taking keen interest on quantum computing. In fact, in a pioneering initiative IBM launched cloud based access to quantum computers for free \cite{santos2016ibm}, while  Google has claimed to achieve quantum supremacy \cite{arute2019quantum}. All these successes have made the common people interested about quantum phenomena and technologies. Further, since most of the research is supported by public funding through taxpayers money, so common people ought to know at-least some part of what the researchers are currently exploring. 

As it appears from the above discussion, the basic ideas of quantum technology may not be easy to follow for an outsider without any background of physics. So, it is important to propagate these ideas to a normal person using the terminology that doesn't involve mathematics.  Hence, we plan to use the analogy of candies with some special properties in order to explain the fascinating quantum phenomena that has some important emerging applications. These candies with special features may be called as ``quantum candies" or qandies. In fact, efforts to explain the ideas of quantum mechanics using candies happened earlier too. For example, in 2000 John Miller used candies to elaborate the concepts of quantum mechanics \cite{miller2000m}. Miller used  qandies with a variety of  physical properties to mimic the abstract concept of wave function. Further, they had also used qandies to elaborate a set of important concepts like quantum mechanical operators, eigenfunctions, expectation values, which are commonly used to describe and understand the quantum phenomena. This was a beautiful use of quantum-classical analogy  to explain the concepts of quantum mechanics to undergraduate chemistry students. Arguably, it weakly influenced the development of a computational model for physical chemistry aimed for undergraduate students \cite{hessley2004computational}. 

Though Miller's ideas were interesting, in what follows we will use another type of qandies which was introduced between 2006-09 by Kayla Jacobs who was an undergraduate student of MIT at that time. The work was never published, but it is discussed in detail in later works of Tal Mor and others (cf. Section 2.2 of \cite{lin2020quantum}; also see Section 2.2 of \cite{lin2021illustrating}; for further reference see : \url{https://youtu.be/0YfkcMm2ajg}). Kayla Jacobs' idea was to intuitively explain the well known BB84 quantum key distribution (QKD) protocol \cite{bennett1984quantum} using quantum candies to school kids. The qandies described by Kayla Jocobs were similar to Miller's candies, but these were indistinguishable (as they were initially covered) until some operation is performed. She  considered black box (or machines) producing qandies {which can have}  two different colors and two different tastes. Jacobs correlated the notion of quantum bits to the notion of qandies that have both color and taste but both of them cannot be discovered simultaneously. This is similar to uncertainty relations and such qandies was used to describe the famous BB84 protocol for quantum key distribution (QKD). Recently, it has been realized than Jacobs' idea can be extended to illustrate various ideas related to quantum computing and quantum communication. For example,  in 2020, Lin and Mor \cite{lin2020quantum} introduced the idea of pseudo-entangled qandies and correlated qandies by generalising Jacobs' idea to two pair of qandies. This generalization made it possible to go beyond BB84 protocol and to explain a large number of other protocols for QKD. Specifically, they used the qandy pairs to illustrate a set of protocols for QKD. They actually introduced correlation measuring device for correlated qandies in order to illustrate a set of protocols for quantum cryptography that requires correlated qubits (entangled states or Bell nonlocal states) including  measurement device independent (MDI) protocol for QKD. Further, in 2021, Lin and Mor has successfully illustrated the concept of quantum bit commitment using qandies \cite{lin2021illustrating}. Moreover, the idea of quantum digital signatures has also been recently explained using qandies  \cite{mor2022digital}.  All these works have established that qandies can serve as an important tool that allows us to share the exciting development in the domain of quantum computing and communication with the general public. However, until now only a few interesting concepts (mentioned above) have been explained using qandies with many other interesting ideas including quantum teleportation which are yet to be explained in such simple terminology. Motivated by this fact, here we aim to illustrate the idea of quantum teleportation using qandies. The main idea of quantum teleportation was given by Bennett et al. \cite{bennett1993teleporting}. The teleportation protocol is successful when the communicating parties share an entanglement and the classical result of measurement. Constructing similar analogy with quantum candies, pseudo-entangled qandies and correlated qandy measure introduced, we depict qandy teleportation protocol for a single qandy. We mainly focus on illustrating how pseudo-entangled qandies help the communicating parties Alice and Bob in teleportation and how unknown qandy is teleported at receivers end. The teleportation protocol presented here does not include any mathematical calculations, but the quantum mechanical principles. 
 
The paper is organised as follows. In section \ref{sec-1}, we introduce the idea of quantum candies and the quantum mechanical principles behind it. In Section \ref{sec-2}, we describe the teleportation protocol of Bennett et al. using the newly introduced  quantum candies. Then, in Section \ref{sec-3} we summarize our results and conclude the paper.

\section{Quantum candies}\label{sec-1}

In this section, we present Kayla Jacobs' model of quantum candies in which she introduced quantumness in qandies to develop the idea of quantum uncertainty principle to teach quantum key distribution in a simpler way. Jacobs considered qandies with two general properties i.e., color and taste. She set qandies colors \{Red (R), Green (G)\} and tastes \{Vanilla (V), Chocolate (C)\}, which is equivalent to forming two basis sets. These general properties of color and taste {can be considered}  for describing the phenomenon of quantumness present in qandies. 

\begin{figure}[]
\centering
\includegraphics[width=1.0\linewidth]{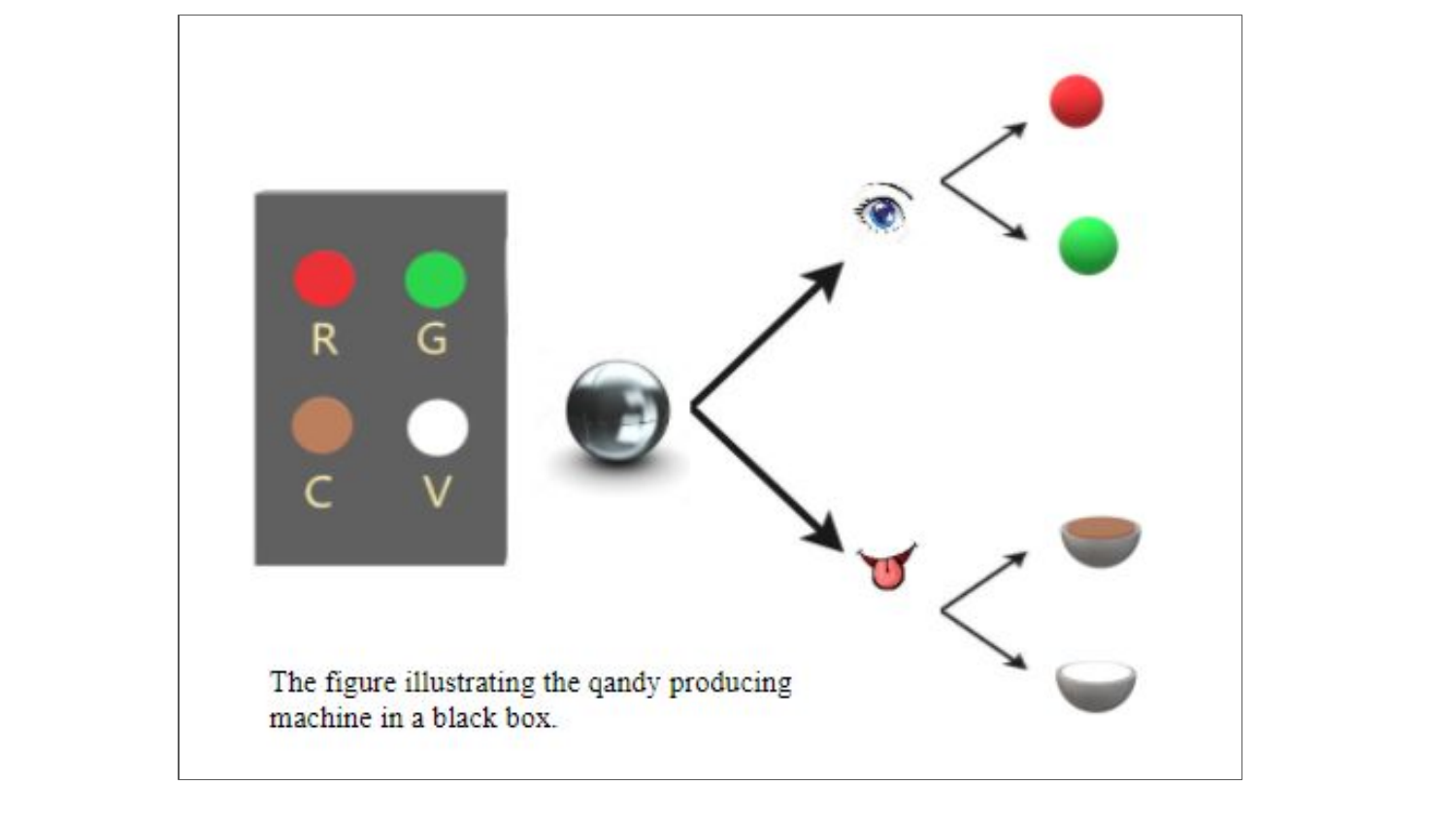}
\caption{(Color Online). Brief description on the quantumness of a single qandy.}
\label{figure-1}
\end{figure}

In Fig.\ref{figure-1}, we {pictorially represent}  the Jacobs' model of quantum candies and  thus illustrate the {quantum features}. These qandies have two general properties, color and taste (as the classical candies).  Further, the qandies coming out of the machine are wrapped by some paper. The qandy machine  has four buttons and one can only choose one of the specific properties \{R, G, V, C\} at a time. Once a color is chosen, the taste is random, {i.e.} if someone decides to eat a red candy, then the taste \{V, C\} is random, but if someone desires to look then the color \{R \} would be correct. Similarly, if a specific taste of a qandy is chosen, the color  \{R, G\} is random, but the taste \{V, C\} would be correct. It can be seen that  machine produces a qandy which has a general property { in which at a particular instant one cannot correctly guess both the taste and color.} One has to perform an operation of look or taste to learn about the color (R or G) or about the taste (V or C) of the qandy. It is equivalent to the  form of complementary principle in quantum physics {with both the color and taste  not being defined together.} 

For illustrating BB84 protocol \cite{bennett1984quantum}, Kayla Jacobs replaced sending qubits by qandies. Here, the classical bits is mapped to color and taste. Now, any person looking at a qandy has four options which has one to one correspondence as quantum bits and is written as \{\{{0, 0}\} ←→ \{C, R\}[taste,color],  \{0, 1\} ←→ \{C, G\}[taste,color], \{1, 0\} ←→ \{V, R\}[taste,color] and \{1, 1\} ←→ \{V, G\}[taste,color]\}. The qandies with two different colors and tastes in the set form an analogy to computational basis in 2-dimensional Hilbert space. For BB84 protocol, Alice randomly prepares a string of qandies either in {color or taste} state and sends it to Bob. Bob then decides to either look or taste the qandies sent by Alice and then making a note of the outcome. In this way, both Alice and Bob are able to share the information about color and taste and hence are able to generate a key. For example, if Alice sends Red qandy, Bob looks at it, then it is R. But if there is some Eve who wants know about the qandy, then she measures the qandy by looking at it. Eve then prepares the same new qandy and sends it to Bob. Here, Eve is lucky and gets full information. But if Eve eats the qandy, she measures a random taste \{C or V\}, and gets no information. Similarly, one could define qandies for more than two general properties for further demonstrations of quantum physics. Jacobs qandies model was extended by Lin-Mor in producing a pair of qandies by a new type of qandy machine. The machine generates a pair of qandies in unknown colors and tastes with correlations resembling entangled states. Further the model was exended to Lin-Mor-Shapira qandies by producing a machine that distinguishes various correlated qandy states. Moreover, our current protocol of qandy teleportation relies on such qandy producing machines, and is described in the next section.     

\section{Illustrating the qandy teleportation}\label{sec-2}

In this section, we describe the teleportation protocol using the model developed by Lin–Mor and Lin–Mor–Shapira for qandies. Imagine two parties, Alice and Bob who wish to communicate to each other in this protocol. Here, Alice is the sender and Bob is the recipient. We shall see how qandies are teleported and how non-classical properties of qandies assist the communicating parties to accomplish the task. In order to exploit qandy teleportation, we consider the qandies with single-qandy state, multi-qandy state, qandies-correlations (pseudo-entanglement) and qandy measurements. Also, we have some qandy gates which perform some qandy operation with respect to input qandies. These qandies react differently to the a given sigle gate.

To begin with, we describe the correlated qandies developed by Lin–Mor. They descibed a new type of qandy producing machine that generates four types of correlated qandy pairs and is denoted by \{\{$\phi_+$\}, \{$\psi_+$\}, \{$\phi_+$\}, \{$\psi_-$\}\}. The correlated qandies has a property that, if they are measured (either looked or tasted) they give random outcome, but correlated. It can be described as:
\begin{itemize}
    \item \{$\phi_+$\}: the outcome of the correlated qandies can be random, but gives identical outcomes of taste or look (RR or GG or VV or CC);
    \item \{$\psi_+$\}: If both candies are looked at, they represent opposite colors (RG or GR) but when it's tasted, it resembles similar flavor (VV or CC) i.e, they present identical tastes;
    \item \{$\phi_-$\}: the correlated qandies resemble same color when looked (RR or GG) but tastes different when they are tasted (VC or CV); 
    \item \{$\psi_-$\}: here both the qandies when they are looked has opposite colors or when they are tasted has the different tastes (RG or GR or VC or CV ).
\end{itemize}
These correlated qandy pairs acts as pseudo-entangled qandies and can be used for teleportation. These characteristic properties of qandies mimic the quantum entanglement. The correlated qandies are different from single qandy described in previous section. For example, in \{$\phi_+$\} if both the qandies are looked at, the oucome would be (RR or GG) each occuring with a probability of 1/2; similarly, if they are tasted, the oucome would be (VV or CC) with a probability of 1/2. If one qandy is looked and other is tasted, the measurement would give a random outcome, and there exists no correlations between the two outcomes. Hence correlated-qandies are used for sharing pseudo-entanglement between communicating parties in any QKD protocol.\\

Now, let us describe Lin–Mor–Shapira's qandy “Pseudo-Bell measurements”. It is a machine (or black box) that distinguishes all the four correlated qandy pairs without looking or tasting at it. The machine measurements can be defined as: 
\begin{itemize}
    \item If the two correlated qandies are produced with identical colors (RR or GG), the machine observes $\{\phi+\}$ or $\{\phi-\}$, with same probabilities.
    \item If the two correlated qandies are prepared with identical tastes (VV or CC), the machine observes $\{\phi+\}$ or $\{\psi+\}$, with same probabilities.
    \item If the two correlated qandies are prepared with opposite colors (RG or GR), the machine observes $\{\psi+\}$ or $\{\psi-\}$, with same probabilities.
    \item If the two correlated qandies are prepared with opposite tastes (CV or VC), the machine will observe $\{\phi-\}$ or $\{\psi-\}$, with same probabilities.
    \item If one qandy is prepared in a  definite color and the other with a definite taste, all the four correlated qandy states will be observed with equal probabilities.
\end{itemize}
In quantum information, quantum gates are the basic units describing the transformations. Likewise, in the qandy world, qandy gates are defined as applying the similar operations irrespective of the inputs, and different qandies react differently to the same gate. The simple qandy gates on single qandies can be defined as color-switching gate and the taste-switching gate representing the Pauli X gate and Pauli Z gate in quantum computing. The gate operations of the single qandy color-switching and the taste-switching gates are defined in Table 1. Similarly, the CNOT gate operation and the gate operations for the correlated qandies are defined.
\begin{table}[ht]\label{table-1}
\caption{Input-output conditions for a qandy X (color-switching) and qandy Z (tasteswitching) gates}
\centering 
\begin{tabular}{c c c c}
\hline\hline 
Input & X & Z\\ [0.5ex]
\hline                 
\{R\} & \{G\} & \{R\} \\
\{G\} & \{R\} & \{G\}  \\
\{C\} & \{C\} & \{V\}   \\
\{V\} & \{V\} & \{C\}  \\[1ex]
\end{tabular}
\end{table}

\begin{figure}[]
\centering
\includegraphics[width=1.00\linewidth]{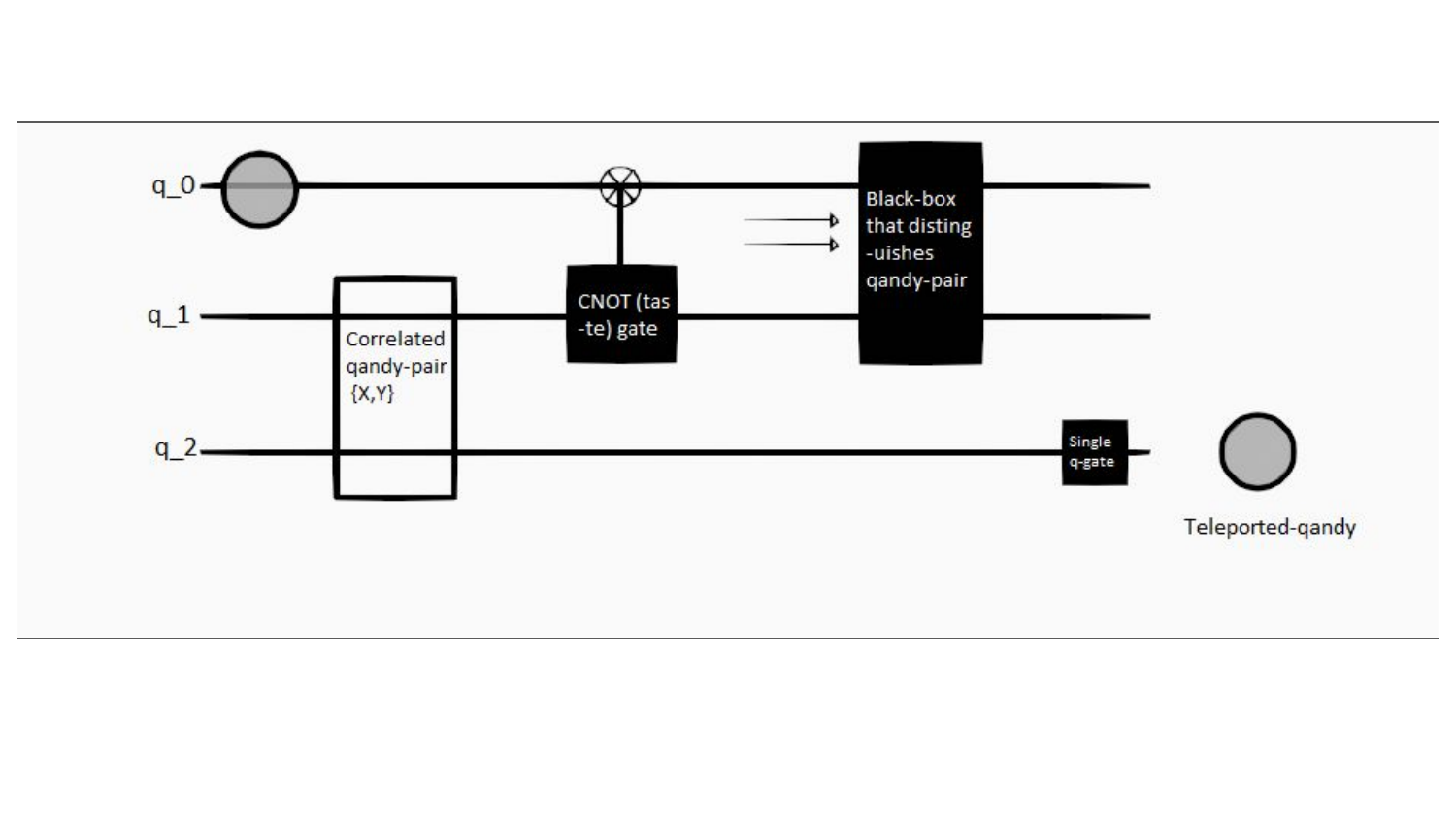}
\caption{(Color Online). The qandy circuit representing teleportation of a single qandy. Here the state has $q\_0$ is with Alice, and the states $q\_1, q\_2$ are with Alice and Bob. The qandy pair is denoted by \{X, Y\}. Then CNOT gate is applied on qandies $q\_0, q\_1$ and in the next step black box is placed for measurement.}
\label{figure-2}
\end{figure}

Using all the properties developed, we will now construct the manual for qandy teleportation. Fig. \ref{figure-2} illustrates the single qandy teleportation between the two communicating parties Alice and Bob who are spatially separated. The teleportation protocol is successful if the quantum state is telported from Alice to Bob via use of pre-shared entanglement and local operations and classical communications (LOCCs). Similar exercise will be now done by use of qandies. In qandy teleportation, initially an observer Alice will have a single qandy which is represented by a small spherical object on $q\_0$ and the channels $q\_1, q\_2$ are shared between Alice and Bob is seen in Fig. \ref{figure-2}. At this point, Alice has no information about the color or taste of qandy on $q\_0$. Let's assume the general property of qandy here is taste (for simplicity, the taste of unknown qandy be chocolate \{C\} and it is unknown to Alice) and she wishes to send information of the single qandy to Bob. The pseudo-entangled correlated qandy pair shared between observers be $\{\phi_+\}$ (i.e., the outcome of the correlated qandies can be random but identical. We consider here the taste outcome of qandies). The correlated qandy pair shared between observers is given as $\{V\}_1\{V\}_2$. Each of the observers Alice and Bob gets one qandy which has correlation property. Let Alice's qandy be $\{V\}_1$ and Bob's qandy be $\{V\}_2$. Now, Alice can perform any local operations of look or taste on her part on $q\_0, q\_1$ (or on her qandies \{C\} and $\{V\}_1$). Alice tries to control both her qandies by performing a CNOT taste-gate operation. The action of the qandy CNOT (taste-gate) on correlated qandies is defined as, the gate operation that switches the taste of second qandy (target-qandy) if the first qandy (control-qandy) is in \{V\} state. For example, if the correlated pair is given by $\{C\}_1\{C\}_2$, the gate operation does nothing and returns $\{C\}_1\{C\}_2$ (as the control qandy is $\{C\}_1$), even if the second qandy $\{C\}_2$ has any property of color or taste. Instead, if the qandy pair is given by $\{V\}_1\{C\}_2$, the CNOT gate operation transfers it to $\{V\}_1\{V\}_2$ (since the control qandy is $\{V\}_1$). For further information about CNOT taste gate refer to \cite{lin2021illustrating}.

Now, Alice tries to perform CNOT taste gate operation on her qandies by sending it to a blackbox. Here, Alice's unknown qandy's taste operation will be performed and is given as $\{C\}_1\{V\}_2$ and then further it is to be sent for measurement. Alice now performs Lin–Mor–Shapira’s qandy “Pseudo-Bell measurements" by sending her qandies to black box that perfectly distinguishes the correlated qandy pair without looking or tasting at it. Since, the two correlated qandies tastes are opposite the machine will observe $\{\phi_-\}$ or $\{\psi_-\}$, with equal probabilities. The oucome of the Alice's state will have, different tastes when they are tasted. Alice now communicates about her outcome (look/taste) to Bob via the classical communication. Bob, accordingly performs (look/taste) to get information about unknown qandy. For example, if Alice get's her outcome as $\{C\}_1\{V\}_2$, Bob just tastes the qandy and obtains \{C\} on $q\_2$. If Alice's oucome is $\{V\}_1\{C\}_2$, Bob need's to perform taste-switching operation on his part $q\_2$ and get's \{C\} depending on the classical communication from Alice. This way, on Bob's state $q\_2$ the information about the unknown qandy \{C\} is teleported. Thus, based on general properties (like look or taste) of qandies, the teleportation protocol is performed faithfully. One can also change the physical properties of qandies or increase the number of sub-properties of taste/color (include 4 colors or tastes) to learn about various protocols in quantum communication processing.

\section{Conclusion}\label{sec-3}
The study has introduced the fundamental properties of quantum information theory in terms of newly introduced model ``quantum candies (qandies)". These qandies provide a lucid way for understanding the concepts of quantum information for people who do not belong to this area of research but are interested to know the basics for their understanding. So, in order to explain the quantum teleportation to a larger audience, we have correlated the correspondence of quantum world to the domain of qandy world and successfully accomplished the teleportation of the qandy. This study has been able to explain the features of quantum information to the students and general public without the use of  mathematics. This model will help the science and engineering educators to provide a mechanism to propagate the deeper concepts of quantum science and technology to the general public.  

\section{Acknowledgments}
NN and AP acknowledges the support from Interdisciplinary Cyber Physical Systems (ICPS) programme of the Department of Science and Technology (DST), India, Grant No.: DST/ICPS/QuST/Theme-1/2019/6 (Q46).

\bibliographystyle{ieeetr}
\bibliography{qandies}

\begin{thebibliography}{10}

\bibitem{dowling2003quantum}
J.~P. Dowling and G.~J. Milburn, ``Quantum technology: the second quantum
  revolution,'' {\em Philosophical Transactions of the Royal Society of London.
  Series A: Mathematical, Physical and Engineering Sciences}, vol.~361,
  no.~1809, pp.~1655--1674, 2003.

\bibitem{ladd2010quantum}
T.~D. Ladd, F.~Jelezko, R.~Laflamme, Y.~Nakamura, C.~Monroe, and J.~L.
  O’Brien, ``Quantum computers,'' {\em nature}, vol.~464, no.~7285,
  pp.~45--53, 2010.

\bibitem{pirandola2020advances}
S.~Pirandola, U.~L. Andersen, L.~Banchi, M.~Berta, D.~Bunandar, R.~Colbeck,
  D.~Englund, T.~Gehring, C.~Lupo, C.~Ottaviani, {\em et~al.}, ``Advances in
  quantum cryptography,'' {\em Advances in optics and photonics}, vol.~12,
  no.~4, pp.~1012--1236, 2020.

\bibitem{mannalatha2023comprehensive}
V.~Mannalatha, S.~Mishra, and A.~Pathak, ``A comprehensive review of quantum
  random number generators: Concepts, classification and the origin of
  randomness,'' {\em Quantum Information Processing}, vol.~22, no.~12, p.~439,
  2023.

\bibitem{bennett1993teleporting}
C.~H. Bennett, G.~Brassard, C.~Cr{\'e}peau, R.~Jozsa, A.~Peres, and W.~K.
  Wootters, ``Teleporting an unknown quantum state via dual classical and
  einstein-podolsky-rosen channels,'' {\em Physical review letters}, vol.~70,
  no.~13, p.~1895, 1993.

\bibitem{hu2023progress}
X.-M. Hu, Y.~Guo, B.-H. Liu, C.-F. Li, and G.-C. Guo, ``Progress in quantum
  teleportation,'' {\em Nature Reviews Physics}, vol.~5, no.~6, pp.~339--353,
  2023.

\bibitem{wootters1982single}
W.~K. Wootters and W.~H. Zurek, ``A single quantum cannot be cloned,'' {\em
  Nature}, vol.~299, no.~5886, pp.~802--803, 1982.

\bibitem{lin2020quantum}
J.~Lin and T.~Mor, ``Quantum candies and quantum cryptography,'' in {\em
  International Conference on the Theory and Practice of Natural Computing},
  pp.~69--81, Springer, 2020.

\bibitem{lin2021illustrating}
J.~Lin, T.~Mor, and R.~Shapira, ``Illustrating quantum information with quantum
  candies,'' {\em arXiv preprint arXiv:2110.01402}, 2021.

\bibitem{mor2022digital}
T.~Mor, R.~Shapira, and G.~Shemesh, ``Digital signatures with quantum
  candies,'' {\em Entropy}, vol.~24, no.~2, p.~207, 2022.

\bibitem{feynman1986quantum}
R.~P. Feynman, ``Quantum mechanical computers.,'' {\em Found. Phys.}, vol.~16,
  no.~6, pp.~507--532, 1986.

\bibitem{santos2016ibm}
A.~C. Santos, ``The {IBM} quantum computer and the {IBM} quantum experience,''
  {\em arXiv preprint arXiv:1610.06980}, 2016.

\bibitem{arute2019quantum}
F.~Arute, K.~Arya, R.~Babbush, D.~Bacon, J.~C. Bardin, R.~Barends, R.~Biswas,
  S.~Boixo, F.~G. Brandao, D.~A. Buell, {\em et~al.}, ``Quantum supremacy using
  a programmable superconducting processor,'' {\em Nature}, vol.~574, no.~7779,
  pp.~505--510, 2019.

\bibitem{miller2000m}
J.~B. Miller, ``The {M\&M} superposition principle,'' {\em Journal of Chemical
  Education}, vol.~77, no.~7, p.~879, 2000.

\bibitem{hessley2004computational}
R.~K. Hessley, ``A computational-modeling course for undergraduate students in
  chemical technology,'' {\em Journal of Chemical Education}, vol.~81, no.~8,
  p.~1140, 2004.

\bibitem{bennett1984quantum}
C.~H. Bennett and G.~Brassard, ``Quantum cryptography: {P}ublic key
  distribution and coin tossing,'' in {\em International Conference on Computer
  System and Signal Processing, IEEE, 1984}, pp.~175--179, 1984.

\end{thebibliography}

\end{document}